\newcommand{\PRE}[1]{}      
\documentclass[aps,twocolumn,amsmath,preprintnumbers,amsfonts,amssymb]{revtex4}
\usepackage{bm}
\usepackage{epsfig}
\usepackage{graphicx}

\newcommand{\be}{\begin{equation}}
\newcommand{\ee}{\end{equation}}
\newcommand{\bea}{\begin{eqnarray}}
\newcommand{\eea}{\end{eqnarray}}
\newcommand{\nn}{\nonumber}
\newcommand{\ba}{\begin{array}} 
\newcommand{\ea}{\end{array}}
\newcommand{\lmat}[2][lllll]{\left( \begin{array}{#1} #2\\ \end{array} \right)}

\newcommand{\rmat}[2][rrrrr]{\left( \begin{array}{#1} #2\\ \end{array} \right)}

\newcommand{\lsim}{
\mathrel{\hbox{\rlap{\hbox{\lower4pt\hbox{$\sim$}}}\hbox{$<$}}}}
\newcommand{\gsim}{
\mathrel{\hbox{\rlap{\hbox{\lower4pt\hbox{$\sim$}}}\hbox{$>$}}}}
\newcommand{\til}{\widetilde}

\newcommand{\tev}{\text{TeV}}
\newcommand{\gev}{\text{GeV}}

\begin{document}

\preprint{UCRHEP-T460}

\title{
\PRE{\vspace*{1.5in}}
Dileptons and four leptons at $\boldsymbol{Z'}$ resonance in the early stage of the LHC
\PRE{\vspace*{0.3in}} }
\author{Hye-Sung Lee}
\affiliation{
Department of Physics and Astronomy, University of California, Riverside, CA 92521, USA
\PRE{\vspace*{.5in}}
}

\begin{abstract}
\PRE{\vspace*{.1in}} \noindent
The LHC era just began.
The first discovery at the LHC experiment would be arguably a new resonance pole at TeV scale, if it exists.
While the discovery of the $Z'$ would be exciting by itself, it may also suggest what other new physics signals should be looked for while the LHC experiment is still at its early stage. 
We argue that the four lepton resonance at the $Z'$ pole is a well-motivated and promising signal especially in supersymmetry framework, which can serve as a supersymmetry search scheme even in the early stage of the LHC experiment.
\end{abstract}


\maketitle

\section{Introduction}
The Large Hadron Collider (LHC) era finally arrived.
Though the LHC aims at many targets such as Higgs boson and supersymmetric particles, the earliest discovery is expected to be a resonance pole at TeV scale, if it exists.
The dilepton resonance would be a clean signature even at the hadron collider, and the spin of the new gauge boson $Z'$ can be easily verified from the angular distribution of the dilepton.

While it could be any neutral component of the $SU(N)$ gauge boson, we consider Abelian gauge group $U(1)'$ for its origin.
The $U(1)'$ is predicted by many new physics scenarios including extra dimensions, grand unified theories, and string theories\footnote{See Refs.~\cite{Hewett:1988xc,Langacker:2008yv} for reviews on the $U(1)'$.}.
More recently, it has appeared in the hidden valley models \cite{Strassler:2006im}.

Though $Z'$ is considered as a source of the fermion pair in most experimental setup, it can couple to the gauge boson pair and the scalar pair as well.
The gauge boson case, for example, $W^+W^-$ pair is possible when there is a mixing between $Z$ and $Z'$ \cite{delAguila:1986ad,Barger:1987xw,Deshpande:1988py}.
The scalar case can be a Higgs or other kind of scalar.
The $Z'$ decaying to the Higgs pair can possibly serve as a good channel for Higgs search.
It can be probed by looking for heavy particles such as the electroweak gauge bosons or heavy fermions.

There may be other kind of scalars, in general, whose major decay mode is a clean signal of light charged leptons ($e$, $\mu$).
Sfermions are the scalars naturally provided in the supersymmetry (SUSY) framework, as superpartners of the fermions.
Sfermions would decay to the lightest supersymmetric particle (LSP) through a cascade decay.
If a sfermion is the LSP, it may still decay to the standard model (SM) fermions if the $R$-parity is absent.
In contrast to the Higgs case where the dominant decay mode is related to the masses of final particles, the scalar LSP decay through the $R$-parity violating coupling may have the light fermions as a major decay channel.
Superpartner pairs may be produced abundantly by the new resonance \cite{Gherghetta:1996yr,Baumgart:2006pa}.
The sfermion LSP pair can decay to 4 fermions making a resonance at $Z'$ pole in the absence of the $R$-parity.

\begin{figure}[tb]
\includegraphics[width=0.3\textwidth]{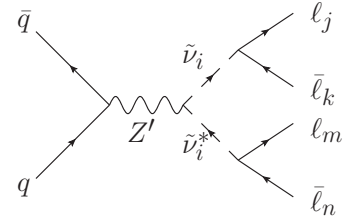}
\caption{4 lepton resonance through $Z'$ and sneutrino pair.}
\label{fig:4leptons}
\end{figure}

If the LSP is the sneutrino ($\til \nu$), the lepton number violating term $\lambda LLE^c$ would give 4 charged lepton resonance, which could be a clean signal even in the early stage of the LHC experiment (see Figure \ref{fig:4leptons}).
Therefore, the ``$Z' \to$ scalar pair $\to 4$ leptons'' mode is not only a novel channel for the LHC but also a natural scenario motivated by the SUSY.

We consider this lepton number violating model with the $\til \nu$ LSP as our example to study the feasibility of the 4 lepton channel at the LHC.
In general, it applies to any new physics scenario that has the $Z'$ coupling to scalar and the scalar coupling to charged leptons though.
In the SUSY case, the charged slepton cannot be too much heavier than the sneutrino, and it would serve as an additional source of the leptons.
This additional contribution will not be included in our numerical analysis since the sneutrino part result alone is large enough to give evidence to support our conclusion.
Since the charged sleptons can only increase the number of leptons, our result can still serve as a minimally expected signal.
If the 4 lepton signals can be discovered early enough, it will serve as a new SUSY signal search scheme.

\section{TeV scale $U(1)'$ gauge symmetry}
A great motivation for the TeV scale extra Abelian gauge symmetry or $U(1)'$ can be found in supersymmetrization of the SM.
The general superpotential of the minimal supersymmetric extension of the SM (MSSM) before $R$-parity is imposed is given as following.
\bea
W &=& \mu H_u H_d \nn \\
&+& y_E H_d L E^c + y_D H_d Q D^c + y_U H_u Q U^c \nn \\
&+& \lambda LLE^c + \lambda' LQD^c + \mu' H_uL + \lambda'' U^cD^cD^c \nn \\
&+& \frac{\eta_1}{M} QQQL + \frac{\eta_2}{M} U^cU^cD^cE^c + \cdots
\eea

SUSY is, arguably, the best motivated new physics paradigm that can address various problems of the SM, most notably the gauge hierarchy problem.
However, a mere realization of the supersymmetric SM has some issues that should be addressed.
Among them are
(1) proton decay problem,
(2) dark matter candidate stability problem, and
(3) $\mu$-problem.
SUSY needs a companion mechanism or symmetry that can address these problems.

$R$-parity is the most popular SUSY companion symmetry, and it guarantees the stability of the LSP, providing a good dark matter candidate if a neutral particle such as neutralino happens to be the lightest among the superparticles.
$R$-parity also prevents the proton decay through the renormalizable lepton number ($\cal L$) violating terms ($LLE^c$, $LQD^c$, $H_u L$) or baryon number ($\cal B$) violating term ($U^cD^cD^c$).
For this reason, the MSSM with $R$-parity has been most extensively studied among the supersymmetric models.
Nevertheless, the $R$-parity does not prevent dimension five $\cal L$ and $\cal B$ violating terms ($QQQL$, $U^cU^cD^cE^c$), which still can mediate too fast proton decay \cite{Weinberg:1981wj}, and the $\mu$-problem still needs another solution.
Although $R$-parity may still be a valid SUSY companion symmetry, possibilities are limited, and it suggests us to consider an alternative SUSY companion symmetry.

It turned out that a new TeV scale Abelian gauge symmetry $U(1)'$ is an attractive alternative to the $R$-parity\footnote{See Ref.~\cite{Lee:2008zzl} for a review on this subject.}.
We will consider this $U(1)'$ as the SUSY companion symmetry.
In this letter, after we briefly review how the $U(1)'$ can help with aforementioned problems in the absence of the $R$-parity, we argue that this scenario suggests 4 lepton resonance at the $Z'$ pole as a plausible channel that can be discovered at the LHC, after $Z'$ discovery, even in the early stage of the experiment.

One of the problems of the MSSM is that it does not explain why its new parameter $\mu$ should be of electroweak (EW) scale as the natural EW symmetry breaking requires \cite{Kim:1983dt}.
The $U(1)'$ gauge symmetry can replace the original $\mu$ term ($\mu H_uH_d$) with an effective $\mu$ term ($h SH_uH_d$) with a Higgs singlet $S$ that spontaneously breaks the $U(1)'$ gauge symmetry (see, for examples, \cite{Cvetic:1997ky,Langacker:1998tc}).
The sfermion masses get extra $D$-term contributions of the $U(1)'$ breaking scale, which should not be much larger than TeV scale in order to preserve the solution to the gauge hierarchy problem.
Once the $U(1)'$ is broken at TeV scale, the effective $\mu$ parameter
\be
\mu_\text{eff} = h\left<S\right>
\label{eq:mueff}
\ee
of EW/TeV scale is dynamically generated.

One of the direct implications of the model is the existence of a new gauge boson at TeV scale, which is accessible with the LHC.
The mass of the $Z'$ gauge boson is given by
\be
M_{Z'}^2 = 2 g_{Z'}^2 \left( z[H_u]^2 \left<H_u\right>^2 + z[H_d]^2 \left<H_d\right>^2 + z[S]^2 \left<S\right>^2 \right)
\ee
where $z[H_u]$, $z[H_d]$, $z[S]$ ($ \left<H_u\right>$, $ \left<H_d\right>$, $ \left<S\right>$) are the $U(1)'$ charges (vacuum expectation values) of the Higgs fields $H_u$, $H_d$, and $S$, respectively.
With TeV scale $\left<S\right>$, $Z'$ mass is expected to be at the same scale.
See a recent review \cite{Langacker:2008yv} and references therein for general aspects of the $U(1)'$ including the $U(1)'$ breaking mechanism.

The $R$-parity could be added on top of the $U(1)'$ or even as a discrete subgroup of the $U(1)'$ in its equivalent form of matter parity, in principle.
In this letter, we consider the $U(1)'$ as an alternative of the $R$-parity and consider only the case that $R$-parity is not conserved.
This allows the $\cal L$ or $\cal B$ violating terms, which are in fact one of the most general predictions of the SUSY.
General review of the $R$-parity violation can be found in Ref. \cite{Barbier:2004ez}.

Without the $R$-parity, the proton may still be sufficiently stable even at the higher dimension level due to the $U(1)'$.
The $U(1)'$ can have $B_3$ (baryon triality) \cite{Ibanez:1991pr} in the MSSM sector naturally as its residual discrete symmetry \cite{Lee:2007fw,Lee:2007qx}.
$B_3$ allows renormalizable level $\cal L$ violating terms ($LLE^c$, $LQD^c$, $H_uL$) while forbidding $\cal B$ violating term ($U^cD^cD^c$).
$B_3$ has a selection rule of
\be
\Delta {\cal B} = 3 \times \text{integer}
\ee
and the proton decay ($\Delta {\cal B} = 1$ process) is completely forbidden by this selection rule \cite{Castano:1994ec}.

The LSP can decay without the $R$-parity and it is not a good dark matter candidate anymore in general.
However, the $U(1)'$ may have a new parity (called $U$-parity) as its residual discrete symmetry for the hidden sector, and the lightest $U$-parity particle (LUP) can be a good hidden sector dark matter candidate \cite{Hur:2007ur}.
The discrete symmetries for both the MSSM sector ($B_3$) and the hidden sector ($U_2$) can be originated from the common $U(1)'$ gauge symmetry that solves the $\mu$-problem, which makes the model highly economic \cite{Lee:2008pc,Hur:2008sy}.
\be
U(1)' ~\to~ Z_6^\text{tot} = B_3 \times U_2
\ee
It was also shown that the LUP dark matter can satisfy current experimental constraints from direct detection and relic density \cite{Hur:2007ur}.

So it is quite clear that the $R$-parity violating $U(1)'$-extended supersymmetric model is not only realistic but also highly motivated alternative supersymmetric model to the usual $R$-parity conserving MSSM, which can address all aforementioned problems.

\section{Couplings}
Now the question is how we can distinguish this model from the usual MSSM besides the $Z'$ pole at the LHC.
How do we know if the discovered $Z'$ originated from the $U(1)'$ of the previous section?

One possible way is to connect the $Z'$ with the $\cal L$ violating process.
We consider a chain of process where the $Z'$ decays into the LSP pair, which then decay into the SM particles through the $\cal L$ violating interaction at the LHC experiment.

The $\til \nu$ LSP pair can decay into 4 SM fermions through the $\cal L$ violating terms ($\lambda LLE^c$, $\lambda' LQD^c$).
For the sake of simplicity in the numerical analysis, we assume a hierarchy of $|\lambda'| \ll |\lambda|$.
Then the $\til \nu$ LSP will decay only leptonically, and also we will not have to consider dilepton production through $s$-channel sneutrinos or $t$-channel squarks which all need sizable $\lambda'$ couplings at the LHC.

If we see both dilepton resonance and 4 lepton resonance at the same invariant mass, it will strongly hint that the $R$-parity is violated and the $U(1)'$ is acting as a SUSY companion symmetry.
This is a novel channel to produce 4 fermions, and we even know where to look since the dilepton resonance will tell us the mass and width of the $Z'$.

We do not assume any right-handed neutrino or sneutrino, and take our sneutrino LSP a pure left-handed one.
The active neutrino can still get mass through the $\cal L$ violating couplings without any right-handed neutrino \cite{Hall:1983id,Grossman:1998py}.

The partial decay width of the $\til \nu$ into dileptons with $\lambda_{ijk} L_i L_j E^c_k$ is given by
\be
\Gamma (\til \nu_i \to \ell_j \bar \ell_k) = \frac{1}{16 \pi} \lambda^2_{ijk} m_{\til \nu_i} .
\ee
The $SU(2)_L$ gauge invariance requires $\lambda_{ijk} = - \lambda_{jik}$, which results in 9 independent parameters in $\lambda_{ijk}$.

The ratio among the 4 light charged lepton final states from the $\til \nu$ LSP pair, for universal $\lambda$ coupling, is given in Table~\ref{tab:decay} when signs of charge are ignored.
For sneutrino mass of a few $\times ~100 ~\gev$, universal $|\lambda|$ coupling as large as ${\cal O}(|\lambda|) \sim 10^{-3}$ is allowed by the lepton flavor violation constraint (such as $\mu \to eee$) \cite{Barbier:2004ez}.
Since the $\lambda$ coupling is the only channel the $\til \nu$ LSP can decay through, the exact value of $\lambda$ is not relevant unless it is too small, causing a displaced vertex.
Taking into account of the $\tau$ decay into the light leptons with additional neutrinos as well as nonuniform $\lambda$ may alter the ratio.

\begin{table}[tb]
\begin{center}
\begin{tabular}{|l|ccccc|c|}
\hline
LSP pair               & $eeee$ & $eee\mu$ & $ee\mu\mu$ & $e\mu\mu\mu$ & $\mu\mu\mu\mu$ & Br($4\ell$) \\
\hline
~~$\til \nu_e \til \nu_e^*$       & 0   & 0   & 1   & 2 & 1 &  4/36\\
~~$\til \nu_\mu \til \nu_\mu^*$   & 1   & 2   & 1   & 0 & 0 &  4/36\\
~~$\til \nu_\tau \til \nu_\tau^*$ & 1   & 4   & 6   & 4 & 1 & 16/36\\
\hline
\end{tabular}
\end{center}
\caption{
Light charged lepton ($\ell = e, \mu$) decay products ratio of the sneutrino LSP pair depending on flavor under the assumption of uniform $|\lambda_{ijk}|$ ($i \ne j$) and $|\lambda| \gg |\lambda'|$.
Signs of charged leptons are neglected in counting.
Unlisted combinations all contain $\tau$.
\label{tab:decay}}
\end{table}

For a quantitative analysis, we need to specify our $Z'$ couplings for the SM fermions\footnote{The methods to distinguish models at a hadron collider can be found in, for examples, Refs.~\cite{Barger:1986hd,Petriello:2008zr,Godfrey:2008vf}.}.
The general $U(1)'$ charges ($z$) for the SM fermions, which has the $B_3$ and the solution to the $\mu$-problem, is given by \cite{Lee:2007fw,Lee:2007qx}
\bea
\lmat{z[q_L]\\ z[u_R]\\ z[d_R]\\ z[\ell_L]\\ z[e_R]} / A
= b \rmat{1\\ 4\\ -2\\ -3\\ -6}
+ \rmat{0\\ -4\\ 1\\ 1\\ 2}
\label{eq:generalSol}
\eea
when we assume no $SU(2)_L$ exotics.
The coefficient $b$ is any real number, which is from the hypercharge shift invariance property of discrete symmetries.
The charges can be normalized arbitrarily with a scale factor $A$.

The ratio of a partial decay width of $Z'$ into the sneutrino to that of the charged lepton, for each flavor, is determined independent of the $U(1)'$ charge assignment as
\be
\frac{\Gamma (Z' \to \til \nu \til \nu^*)}{\Gamma (Z' \to \ell^+ \ell^-)} = \frac{1}{10} \frac{(M_{Z'}^2 - 4 m_{\til \nu}^2)^{3/2}}{M_{Z'}} ,
\label{eq:42leptonsRatio}
\ee
which is about $1/10$ if $m_{\til \nu} \ll M_{Z'}$.
In fact, this ratio always holds independent of model as long as there is a $LLE^c$ term \cite{Matchev}, which provides the relation
\be
2z[\ell_L] - z[e_R] = 0 .
\ee
Therefore the sneutrino pair produced from the $Z'$ decay is expected to be about $10\%$ of the charged lepton pair generically.

\begin{table}[tb]
\begin{center}
\begin{tabular}{|l|cc|ccccc|c|r|}
\hline
    & $b$ & $g_{Z'} A$ & $\frac{z[q_L]}{A}$ & $\frac{z[u_R]}{A}$ & $\frac{z[d_R]}{A}$ & $\frac{z[\ell_L]}{A}$ & $\frac{z[e_R]}{A}$ & $\frac{\Gamma_{Z'}}{M_{Z'}}$ & $M_{Z'}|_\text{min}$ \\
\hline
(i)  & 0.5 & 0.165 & 0.5 & -2 & 0  & -0.5 & -1 & 0.05 & 900~\gev  \\
(ii) & ~1~ & 0.150 & 1   & 0  & -1 & -2   & -4 & 0.05 & 1100~\gev \\
\hline
\end{tabular}
\end{center}
\caption{
Study points used in the numerical analysis.
\label{tab:study}}
\end{table}

We take two study points
\bea
&& (i)~ b=0.5 , \qquad (ii)~ b=1  \nn
\eea
with common $m_{\til \nu} = 200 ~\gev$.

The SM fermion contributions to the $Z'$ width, in the massless fermion limit, is given by
\bea
\Gamma_{\text SM} &\equiv& 3 ( \Gamma_{\nu \bar \nu} + \Gamma_{e^+ e^-} + 3 \Gamma_{u \bar u} + 3 \Gamma_{d \bar d} ) \\
&=& \frac{3 g_{Z'}^2}{8 \pi} A^2 (53-120 b+84 b^2) M_{Z'} .
\eea
There will be additional contributions such as from Higgs and superparticles, which are hard to quantify since they depend on specific spectrums.
It is sufficient for our purpose to assume the total decay width is given by $\Gamma_{Z'} = 1.1 \Gamma_{\text SM}$, and also assume the $\Gamma_{Z'}$ is $5\%$ of the $M_{Z'}$.
For a detail of the supersymmetric contribution to the $Z'$ width, see Ref. \cite{Gherghetta:1996yr}.
The product of the $U(1)'$ coupling constant $g_{Z'}$ and the scale factor $A$ is then determined as shown in Table~\ref{tab:study}.
The table also shows the current experimental bounds on $M_{Z'}$ for our study points from the Tevatron $e^+e^-$ search \cite{CDFdielectron}.
\bea
&&(i)~ b = 0.5 : M_{Z'} \gsim 900 ~\gev , \nn \\
&&(ii)~ b = 1~ : M_{Z'} \gsim 1100 ~\gev . \nn
\eea

\begin{figure}[tb]
\includegraphics[width=0.45\textwidth]{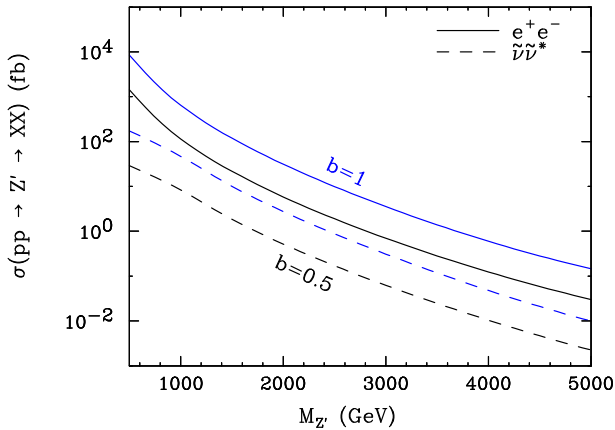}
\caption{Production cross sections for the charged lepton pair (solid) and the sneutrino pair (dashed) for a single flavor at the LHC.}
\label{fig:xsection}
\end{figure}

Figure~\ref{fig:xsection} shows the production cross sections of $p p \to Z' \to \ell^+ \ell^-$ (solid curve) and $p p \to Z' \to \til \nu \til \nu^*$ (dashed curve) for a single flavor of the charged lepton and the sneutrino for our study points in the LHC experiment with $E_{\text CM} = 14 ~\tev$.
For numerical analysis, we use {\tt CompHEP/CalcHEP} \cite{Pukhov:1999gg,Pukhov:2004ca} and the parton distribution function of {\tt CTEQ6L} \cite{Pumplin:2002vw}.
In agreement with Eq.~\eqref{eq:42leptonsRatio}, the sneutrino pair production is slightly smaller than $1/10$ of the dilepton production.

\section{Discovery reach}
Now, we investigate the required luminosity for the 4 lepton events in comparison to the dilepton events.
We use a single lepton flavor for dilepton case, and also a single sneutrino flavor for 4 lepton case.
The production cross section of the dilepton events will be doubled from that in Figure~\ref{fig:xsection} if both electron and muon flavors are counted.
If three generations of the sneutrino are degenerate, the total sneutrino production will be tripled.

For dilepton signal ($p p \to Z' \to \ell^+ \ell^-$), the SM backgrounds are (i) $Z / \gamma^*$ (Drell-Yan), (ii) dijet, and (iii) diboson \cite{Abulencia:2006iv}.
In both dilepton and 4 lepton signals, the $t \bar t$ channel can also give a background, which is reducible \cite{ATLAS}.
The Drell-Yan process will be a dominant background with an appropriate invariant mass cut.

We require the basic cuts on $p_T$, $\eta$, $m_\text{inv}$ as follows.
\begin{itemize}
\item $p_T > 20 ~\gev$ \quad (each lepton)
\item $| \eta | < 2.4$ \quad (each lepton)
\item $| m_\text{inv} - M_{Z'} | < 3 \Gamma_{Z'}$ \quad ($m_\text{inv} = m_{\ell^+\ell^-}$)
\end{itemize}
We find the background is negligible with these cuts for our study points.
Instead of the significance of the signal over background, we just require 10 signal events that pass the cut to claim discovery.

\begin{figure}[tb]
\includegraphics[width=0.45\textwidth]{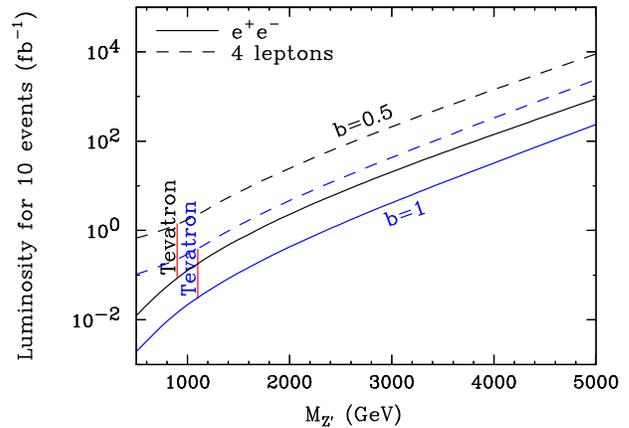}
\caption{Integrated luminosity for 10 events discovery of $Z'$ through dilepton channel, and integrated luminosity for 10 events of 4 light leptons from a single $\til \nu$ pair at the LHC.
The luminosity for 4 lepton events assumes $\text{Br}(4\ell) = 1$.
The vertical lines represent the Tevatron bounds on $Z'$ mass.}
\label{fig:discovery}
\end{figure}

Figure~\ref{fig:discovery} (solid curve) shows the discovery reach of $Z'$ through the dilepton search at the LHC.
For example, the necessary luminosity to discover the resonance by 10 dilepton events for $M_{Z'} = 2 ~\tev$ is (for a single flavor)
\bea
&& (i)~  b=0.5 :~ L = 2.30 ~\text{fb}^{-1} , \nn \\
&& (ii)~ b=1 ~ :~ L = 0.43 ~\text{fb}^{-1} . \nn
\eea
Considering that the first year LHC run is expected to gain total luminosity of about $1 ~\text{fb}^{-1}$, the $Z'$ is expected to be discovered in the early stage of the LHC.

For 4 lepton signal ($p p \to Z' \to \til \nu \til \nu^* \to 4 \ell$), the SM background is $pp \to V V \to 4 \ell$ ($V V = Z Z, \gamma \gamma, \gamma Z$) only with $\ell^+ \ell^- \ell'^+ \ell'^-$ type.
The SM background is small, and furthermore the 4 lepton flavors from the $\til \nu$ LSP pair decay depends on the details sensitively.
For example, it is hard to have $e\mu\mu\mu$ from the SM whose invariant mass can be around the $Z'$ resonance.
The $p p \to H \to Z Z \to 4 \ell$\footnote{Similarly, it is possible to have $p p \to H \to Z' Z' \to 4 \ell$ \cite{Gopalakrishna:2008dv} if $Z'$ is lighter than a half of the Higgs mass.} is not considered as the background here since the Higgs boson is still a new particle we need to search for.

Figure~\ref{fig:discovery} (dashed curve) shows the required integrated luminosity to have 10 events of 4 lepton final states from the $\til \nu$ LSP decay after the same cut with only $m_\text{inv} = m_{4 \ell}$ this time.
For example, the luminosity to have 10 events for $M_{Z'} = 2 ~\tev$ is (for a single $\til \nu$ flavor)
\bea
&& (i)~  b=0.5 :~ L = 25 / \text{Br}(4\ell) ~\text{fb}^{-1} , \nn \\
&& (ii)~ b=1 ~ :~ L = 4.7 / \text{Br}(4\ell) ~\text{fb}^{-1} . \nn
\eea
where $\text{Br}(4 \ell)$ is the branching ratio of a sneutrino pair decaying to 4 light charged leptons.
The 4 leptons can be discovered even in the early stage of the LHC depending on $M_{Z'}$, $\til \nu$ decay branching ratio, and other parameter values.
An individual $\til \nu$ decay branching ratio depends on the flavor of the $\til \nu$ and $\lambda_{ijk}$ (see Table~\ref{tab:decay}, for example).

It is important to keep in mind that these processes can be fully reconstructed and the center-of-mass frame can be found for each event.
Therefore, in principle, one can also measure the spins of $Z'$ and $\til \nu$ easily as well as the $\til \nu$ mass, which would be a discovery of the superparticle.
One way to measure the spin of $\til \nu$ is to use the azimuthal angle between production and decay planes, which was recently advertised in Ref.~\cite{Buckley:2008pp}.
Due to its scalar nature, a flat distribution is expected.

\section{Summary}
In this letter, we suggested a search for the 4 lepton final states at the $Z'$ resonance, with a support from the basic numerical analysis.
Our channel is unique; a spin 1 particle decays into two spin 0 particles and each spin 0 particle decays into charged lepton pair.
This is a well-motivated and attractive channel especially in the SUSY framework when the $R$-parity is replaced by the $U(1)'$ gauge symmetry.
The $U(1)'$ can address the stability of the proton and the dark matter candidate without the $R$-parity \cite{Lee:2008zzl}.
The 4 lepton resonance channel can serve as a new SUSY search scheme even in the early stage of the LHC experiment when the sneutrino is the LSP.
Inclusion of jets and missing transverse energy in the analysis would allow to test scenarios with other types of the LSP as well.

\begin{acknowledgments}
I am indebted to K. Kong and K. Matchev for numerous invaluable discussions.
I am also grateful to P. Konar and S. Rai for useful discussions. 
This work is supported by the US Department of Energy under Grant No. DE-FG03-94ER40837.
\end{acknowledgments}




\begin{thebibliography}{99}
\bibitem{Hewett:1988xc}
  J.~L.~Hewett and T.~G.~Rizzo,
  Phys.\ Rept.\  {\bf 183}, 193 (1989).

\bibitem{Langacker:2008yv}
  P.~Langacker,
  arXiv:0801.1345 [hep-ph].

\bibitem{Strassler:2006im}
  M.~J.~Strassler and K.~M.~Zurek,
  Phys.\ Lett.\  B {\bf 651}, 374 (2007)
  [arXiv:hep-ph/0604261].

\bibitem{delAguila:1986ad}
  F.~del Aguila, M.~Quiros and F.~Zwirner,
  Nucl.\ Phys.\  B {\bf 284}, 530 (1987).
  
\bibitem{Barger:1987xw}
  V.~D.~Barger and K.~Whisnant,
  Phys.\ Rev.\  D {\bf 36}, 3429 (1987).

\bibitem{Deshpande:1988py}
  N.~G.~Deshpande and J.~Trampetic,
  Phys.\ Lett.\  B {\bf 206}, 665 (1988).

\bibitem{Gherghetta:1996yr}
  T.~Gherghetta, T.~A.~Kaeding and G.~L.~Kane,
  Phys.\ Rev.\  D {\bf 57}, 3178 (1998)
  [arXiv:hep-ph/9701343].

\bibitem{Baumgart:2006pa}
  M.~Baumgart, T.~Hartman, C.~Kilic and L.~T.~Wang,
  JHEP {\bf 0711}, 084 (2007)
  [arXiv:hep-ph/0608172].

\bibitem{Weinberg:1981wj}
  S.~Weinberg,
  Phys.\ Rev.\  D {\bf 26}, 287 (1982).

\bibitem{Lee:2008zzl}
  H.~S.~Lee,
  Mod.\ Phys.\ Lett.\  A {\bf 23}, 3271 (2008)
  [arXiv:0811.2539 [hep-ph]].

\bibitem{Kim:1983dt}
  J.~E.~Kim and H.~P.~Nilles,
  Phys.\ Lett.\  B {\bf 138}, 150 (1984).

\bibitem{Cvetic:1997ky}
  M.~Cvetic, D.~A.~Demir, J.~R.~Espinosa, L.~L.~Everett and P.~Langacker,
  Phys.\ Rev.\  D {\bf 56}, 2861 (1997)
  [Erratum-ibid.\  D {\bf 58}, 119905 (1998)]
  [arXiv:hep-ph/9703317].

\bibitem{Langacker:1998tc}
  P.~Langacker and J.~Wang,
  Phys.\ Rev.\  D {\bf 58}, 115010 (1998)
  [arXiv:hep-ph/9804428].

\bibitem{Barbier:2004ez}
  R.~Barbier {\it et al.},
  Phys.\ Rept.\  {\bf 420}, 1 (2005)
  [arXiv:hep-ph/0406039].
  
\bibitem{Ibanez:1991pr}
  L.~E.~Ibanez and G.~G.~Ross,
  Nucl.\ Phys.\  B {\bf 368}, 3 (1992).

\bibitem{Lee:2007fw}
  H.~S.~Lee, K.~T.~Matchev and T.~T.~Wang,
  Phys.\ Rev.\  D {\bf 77}, 015016 (2008)
  [arXiv:0709.0763 [hep-ph]].

\bibitem{Lee:2007qx}
  H.~S.~Lee, C.~Luhn and K.~T.~Matchev,
  JHEP {\bf 0807}, 065 (2008)
  [arXiv:0712.3505 [hep-ph]].

\bibitem{Castano:1994ec}
  D.~J.~Castano and S.~P.~Martin,
  Phys.\ Lett.\  B {\bf 340}, 67 (1994)
  [arXiv:hep-ph/9408230].

\bibitem{Hur:2007ur}
  T.~Hur, H.~S.~Lee and S.~Nasri,
  Phys.\ Rev.\  D {\bf 77}, 015008 (2008)
  [arXiv:0710.2653 [hep-ph]].

\bibitem{Lee:2008pc}
  H.~S.~Lee,
  Phys.\ Lett.\  B {\bf 663}, 255 (2008)
  [arXiv:0802.0506 [hep-ph]].

\bibitem{Hur:2008sy}
  T.~Hur, H.~S.~Lee and C.~Luhn,
  JHEP {\bf 0901}, 081 (2009)
  [arXiv:0811.0812 [hep-ph]].
  
\bibitem{Hall:1983id}
  L.~J.~Hall and M.~Suzuki,
  Nucl.\ Phys.\  B {\bf 231}, 419 (1984).

\bibitem{Grossman:1998py}
  Y.~Grossman and H.~E.~Haber,
  Phys.\ Rev.\  D {\bf 59}, 093008 (1999)
  [arXiv:hep-ph/9810536].
  
\bibitem{Barger:1986hd}
  V.~D.~Barger, N.~G.~Deshpande, J.~L.~Rosner and K.~Whisnant,
  Phys.\ Rev.\  D {\bf 35}, 2893 (1987).

\bibitem{Petriello:2008zr}
  F.~Petriello and S.~Quackenbush,
  Phys.\ Rev.\  D {\bf 77}, 115004 (2008)
  [arXiv:0801.4389 [hep-ph]].

\bibitem{Godfrey:2008vf}
  S.~Godfrey and T.~A.~W.~Martin,
  Phys.\ Rev.\ Lett.\  {\bf 101}, 151803 (2008)
  [arXiv:0807.1080 [hep-ph]].

\bibitem{Matchev}
  K.~T.~Matchev, in private communication.

\bibitem{CDFdielectron}
  [CDF Collaboration],
  CDF/PUB/EXOTIC/PUBLIC/9160.

\bibitem{Pukhov:1999gg}
  A.~Pukhov {\it et al.},
  arXiv:hep-ph/9908288.

\bibitem{Pukhov:2004ca}
  A.~Pukhov,
  arXiv:hep-ph/0412191.

\bibitem{Pumplin:2002vw}
  J.~Pumplin, D.~R.~Stump, J.~Huston, H.~L.~Lai, P.~Nadolsky and W.~K.~Tung,
  JHEP {\bf 0207}, 012 (2002)
  [arXiv:hep-ph/0201195].

\bibitem{Abulencia:2006iv}
  A.~Abulencia {\it et al.}  [CDF Collaboration],
  Phys.\ Rev.\ Lett.\  {\bf 96}, 211801 (2006)
  [arXiv:hep-ex/0602045].

\bibitem{ATLAS}
  [ATLAS Collaboration], ATLAS Technical Design Report, Vol. 2, CERN/LHCC/99-14 (1999).

\bibitem{Gopalakrishna:2008dv}
  S.~Gopalakrishna, S.~Jung and J.~D.~Wells,
  Phys.\ Rev.\  D {\bf 78}, 055002 (2008)
  [arXiv:0801.3456 [hep-ph]].

\bibitem{Buckley:2008pp}
  M.~R.~Buckley, B.~Heinemann, W.~Klemm and H.~Murayama,
  Phys.\ Rev.\  D {\bf 77}, 113017 (2008)
  [arXiv:0804.0476 [hep-ph]].


\end{thebibliography}
\end{document}